\documentstyle[twocolumn,prl,aps,epsf]{revtex}
\begin{document}
\renewcommand{\vec}[1]{{\bf #1}}
\newlength{\figwidth}
\setlength{\figwidth}{0.4 \textwidth}
\addtolength{\figwidth}{-0.5 \columnsep}

\draft

\title{
First Principles Calculation of Elastic Properties of Solid Argon at High Pressures
}
\author{Toshiaki Iitaka\cite{tiitaka}
and Toshikazu Ebisuzaki 
}
\address{
Computational Science Division, \\
RIKEN (The Institute of Physical and Chemical Research) \\
2-1 Hirosawa, Wako, Saitama 351-0198, Japan
}


\maketitle

\begin{abstract}
The density and the elastic stiffness coefficients of fcc solid argon at high pressures from 1 GPa up to 80 GPa are computed by first-principles pseudopotential method with plane-wave basis set and the generalized gradient approximation (GGA). The result is in good agreement with the experimental result recently obtained with the Brillouin spectroscopy by Shimizu et al. [Phys. Rev. Lett. 86, 4568 (2001)].  
The Cauchy condition was found to be strongly violated as in the experimental result, indicating large contribution from non-central many-body force. The present result has made it clear that the standard density functional method with periodic boundary conditions can be successfully applied for calculating elastic properties of rare gas solids at high pressures in contrast to those at low pressures where dispersion forces are important.
\end{abstract}


Rare gas atoms are among the simplest substances in physical and chemical natures because of their closed-shell electronic configuration. Many physical properties of rare gases {\em at low pressures} have been predicted by using  {\it ab initio} or empirical two-body potentials such as the Lennard-Johnes potentials, which consist of short-range repulsive potential and long-range attractive dispersion potential \cite{Aziz1986,Tse1998}.  At low temperatures, the rare gas atoms form van der Waals' crystals of fcc structure except helium, which crystallizes to hcp structure. The crystal structure and the binding energy of rare gas solids have been determined very accurately by experiments. It has been found that two-body potentials can describe these natures very well. Though two-body potentials underestimate the binding energy by few percents, inclusion of three-body potentials as a small correction reduces the error within one percent \cite{Lotrich1997,Rosciszewski2000}.

Recently, the development of the Brillouin spectroscopy in conjunction with diamond-anvil cells (DAC) \cite{Shimizu1981} has opened a door for investigating elastic properties of rare gas solids {\em at high pressures}, which might be important for earth and planetary sciences.
The pressure dependence of the elastic stiffness coefficients has been experimentally determined up to 33 GPa by Grimsditch at al. \cite{Grimsditch1986} and up to 70 GPa by Shimizu et al. \cite{Shimizu2001}.
They used the {\it envelope method} to determine the elastic stiffness coefficients:
The Brillouin frequency shifts are measured without identifying the crystal orientation.
Then the acoustic velocities calculated from the frequency shifts scatter because the acoustic velocity in crystals depends on the propagating direction. 
The maximum and minimum values of acoustic velocities at each pressure are determined and the {\em envelope curves} of the acoustic velocities are drawn. 
The elastic stiffness constants are determined by comparing the envelope curves with the solutions of elastic equation.
Shimizu et al. \cite{Shimizu2001} have also determined the elastic stiffness coefficients of solid argon from 1.6 GPa up to 4 GPa by using {\em in situ} Brillouin spectroscopy \cite{Shimizu1992}, in which the crystal orientation is identified and the more accurate results are obtainable.
From the results of the {\it in situ} Brillouin spectroscopy and the envelope method at high pressures, Shimizu et al. \cite{Shimizu2001} found that solid argon becomes harder than iron \cite{Brazhkin2001} and that the deviation from the Cauchy relation becomes significant. The latter implies that the contribution of non-central many-body force becomes more and more important at higher pressures, and it cannot be treated any more as a small correction to the two-body potentials. 

However, by now, calculations of elastic properties at high pressures have been limited to those with empirical two-body potentials \cite{Tse1998,Grimsditch1986,Occelli2001} as at low pressures. In this paper, therefore, we study the elastic stiffness coefficients of fcc solid argon at high pressures by using first principles calculations with periodic boundary conditions \cite{Kwon1995,McMahan1986,Payne1992} at zero temperature and under constant pressures, which can treat the effects of many-body potentials of crystals in a simple and direct way.
The valence wave functions are expanded in a plane wave basis set truncated at a kinetic energy of 560 eV 
.
The electron-ion interactions are described by the Vanderbilt-type ultrasoft pseudopotentials \cite{Vanderbilt1990}. The effects of exchange-correlation interaction are treated within the generalized gradient approximation of Perdew et al. (GGA-PBE) \cite{Perdew1996}. The model consists of a fcc unit cell containing four argon atoms. The Brillouin zones are sampled with $8 \times 8 \times 8$ Monkhorst-Pack k-points \cite{Monkhorst1976} by using time-reversal symmetry only. During the structural optimization, the enthalpy $H=E+PV$ is minimized by varying the length of the lattice vectors, while the angles between the lattice vectors and the atomic positions in the unit cell are fixed. In the geometrical optimization, the total stress tensor is reduced to the order of 0.001 GPa by using the finite basis-set corrections \cite{Francis1990}.


The elastic stiffness tensor $c_{ijkl}$ relates the stress tensor $\sigma_{ij}$ and strain tensor $\epsilon_{kl}$ by the Hooke's law,
\begin{equation}
\label{eq:Hooke}
\sigma_{ij}=c_{ijkl} \epsilon_{kl}   \mbox{\ \ \ \ \ ($i,j,k,l=x,y,z$)}
.
\end{equation}
Since the stress and strain tensors are symmetric, the most general elastic stiffness tensor has only 21 non-zero independent components. For cubic crystals, they reduce to three components, $c_{11} \equiv c_{xxxx}$, $c_{12} \equiv c_{xxyy}$, and $c_{44} \equiv c_{yzyz}$ (in the Voigt notation).
These elastic stiffness coefficients can be determined by computing the stress generated by forcing a small strain to the optimized unit cell \cite{Wentzcovitch1995,Karki1997}. The lattice vectors $\vec{a}'_i$ of the strained unit cell are determined from the lattice vectors $\vec{a}_i$ of the optimized unit cell by the relation $\vec{a}'_i=(I+\epsilon)\vec{a}_i$, where $I$ is the unit matrix and $\epsilon$ is a strain tensor. 
Two strain tensors,
\begin{eqnarray}
\epsilon^{A} &=& \left(
             \begin{array}{ccc}
              e & 0 & 0 \\ 0 & 0 & 0 \\
             0 & 0 & 0
             \end{array}
             \right)
              \\
\epsilon^{B} &=& \left(
             \begin{array}{ccc}
             0 & e/2 & 0 \\
             e/2 & 0 & 0 \\
             0 & 0 & 0
             \end{array}
             \right) 
,
\end{eqnarray}
are used to determine the three elastic stiffness coefficients, $c_{11}$, $c_{12}$, and $c_{44}$ from Eq.(\ref{eq:Hooke}), namely,
$\sigma_{xx}^A=c_{11} e$, $\sigma_{yy}^A=\sigma_{zz}^A=c_{12} e$ and $\sigma_{xy}^B=c_{44} e$ 
where $\sigma^A$ and $\sigma^B$ are the stress resulting from the strain $\epsilon^A$ and $\epsilon^B$, respectively. 
The $\sigma^A$ and $\sigma^B$ are calculated with $e=0$, $0.01$ and $0.02$ at each pressure, and fitted to a parabolic function of $e$ to remove non-linear contributions. 
The convergence of the elastic stiffness coefficients with respect to the cut-off energy and the number of k-points were estimated of the order of 1 GPa by increasing the cut-off energy to 1120 eV and Monkhorst-Pack k-points to $12 \times 12\times 12$, respectively.

\begin{figure}
\epsfxsize=\figwidth
\epsfbox{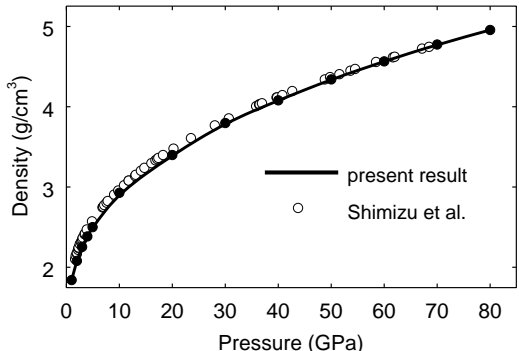}
\caption{
$\rho-P$ equations of state for fcc solid argon. The solid curve with closed circles represents the present result. Open circles represent the experimental result obtained by Shimizu et al. \protect\cite{Shimizu2001} with the Brillouin spectroscopy.
}
\end{figure}

Figure~1 shows the $\rho-P$ equations of state. 
The agreement between the experiment and the GGA calculation indicates that the lattice constant of solid argon  are mainly determined by the balance between the short-range repulsive force and the external pressure. The van der Waals' force, which is not taken into account explicitly in the GGA calculation, is negligible in this pressure range.

\begin{figure}
\epsfxsize=\figwidth
\epsfbox{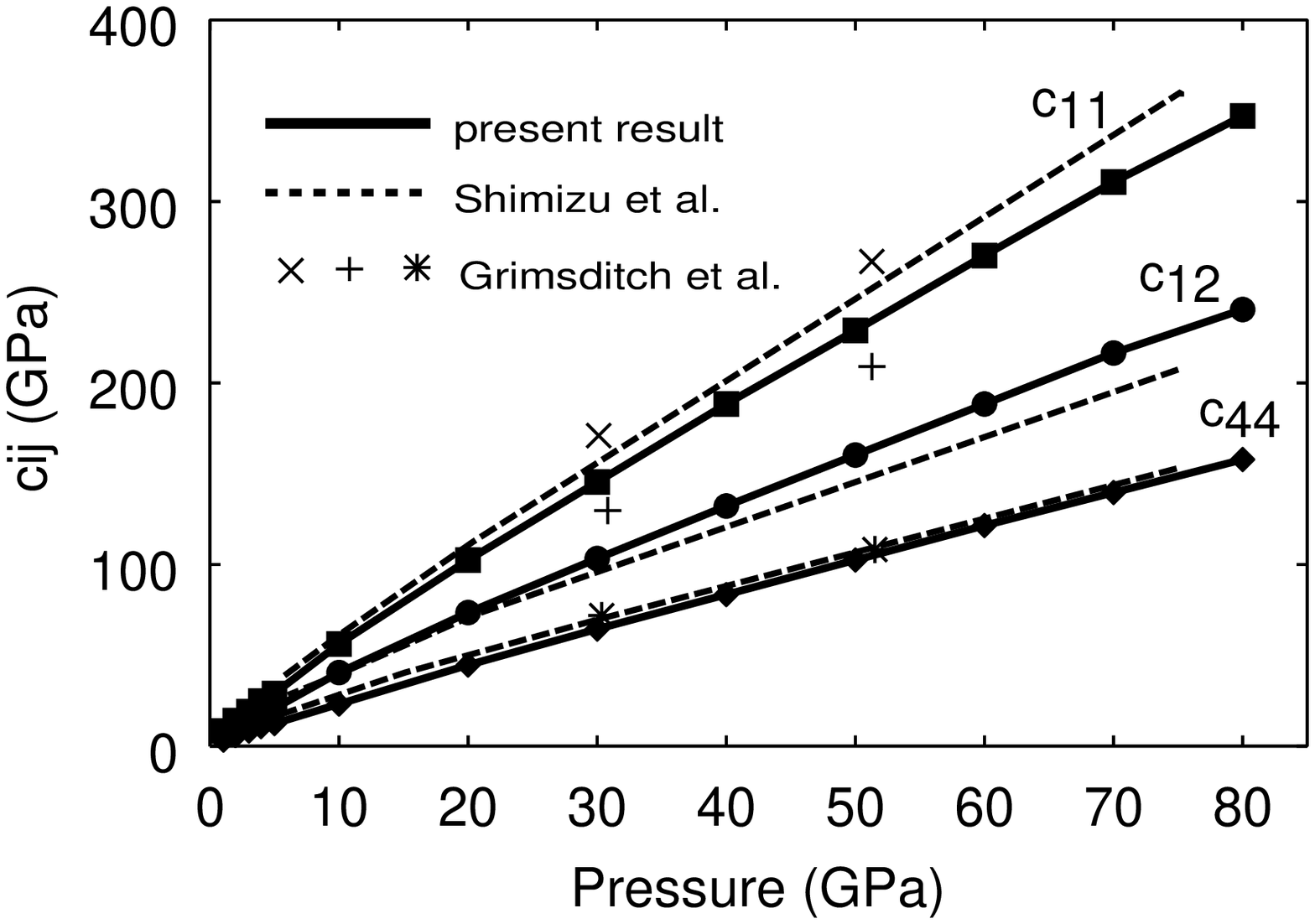}
\epsfxsize=\figwidth
\epsfbox{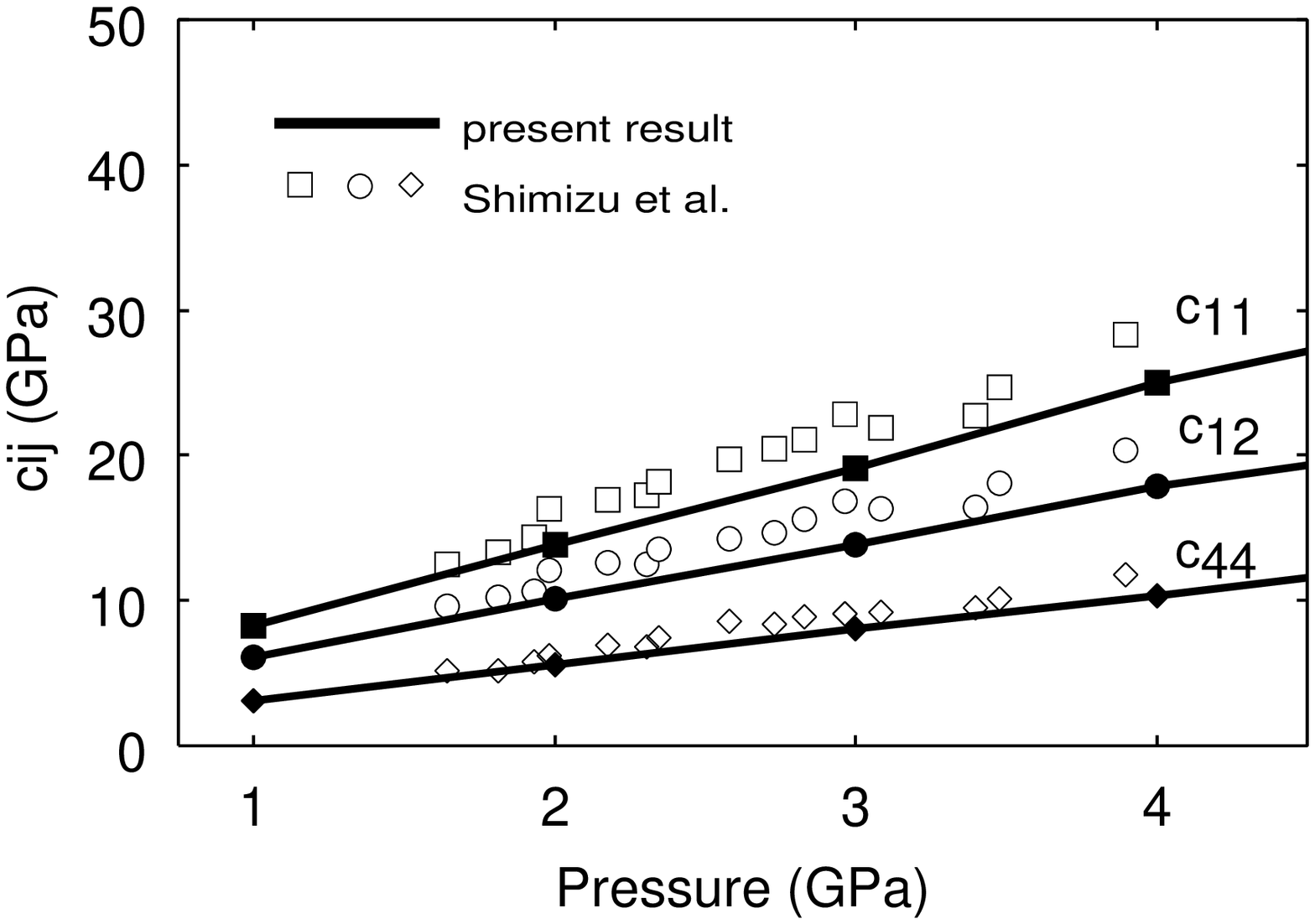}
\caption{ 
The pressure dependence of the elastic stiffness coefficients, $c_{11}$, $c_{12}$, and $c_{44}$, of fcc solid argon. The solid lines with closed symbols represent the present result. The dashed lines represent experimental result at 295 K obtained with envelope method by Shimizu et al. \protect\cite{Shimizu2001}.
Open symbols indicate the result obtained with {\it in situ} Brillouin spectroscopy \protect\cite{Shimizu2001}.
Cross symbols indicate the result of self-consistent phonon calculation based on pair potentials by Grimsditch et al. \protect\cite{Grimsditch1986}.  
}
\end{figure}

Figure~2 shows the pressure dependence of the elastic stiffness coefficients, $c_{11}$, $c_{12}$, and $c_{44}$. 
The elastic stiffness coefficients increase linearly with increasing pressure.
These elastic stiffness coefficients satisfy the generalized elastic stability criteria for cubic crystals under hydrostatic pressure \cite{Wang1993,Wang1995,Karki1997b},
\begin{equation}
c_{11}+2c_{12} > 0, \ \ \ \ \ 
c_{44}>0,           \ \ \ \ \  and \ \ \ \ \ 
c_{11}-c_{12}>0
.
\end{equation}
The agreement between the present and the experimental results for $c_{11}$ and $c_{12}$ does not seem excellent between 10 GPa and 70 GPa. However, the agreement looks better when the acoustic velocities are plotted as a function of the pressure as shown in Fig.~3.


\begin{figure}
\epsfxsize=\figwidth
\epsfbox{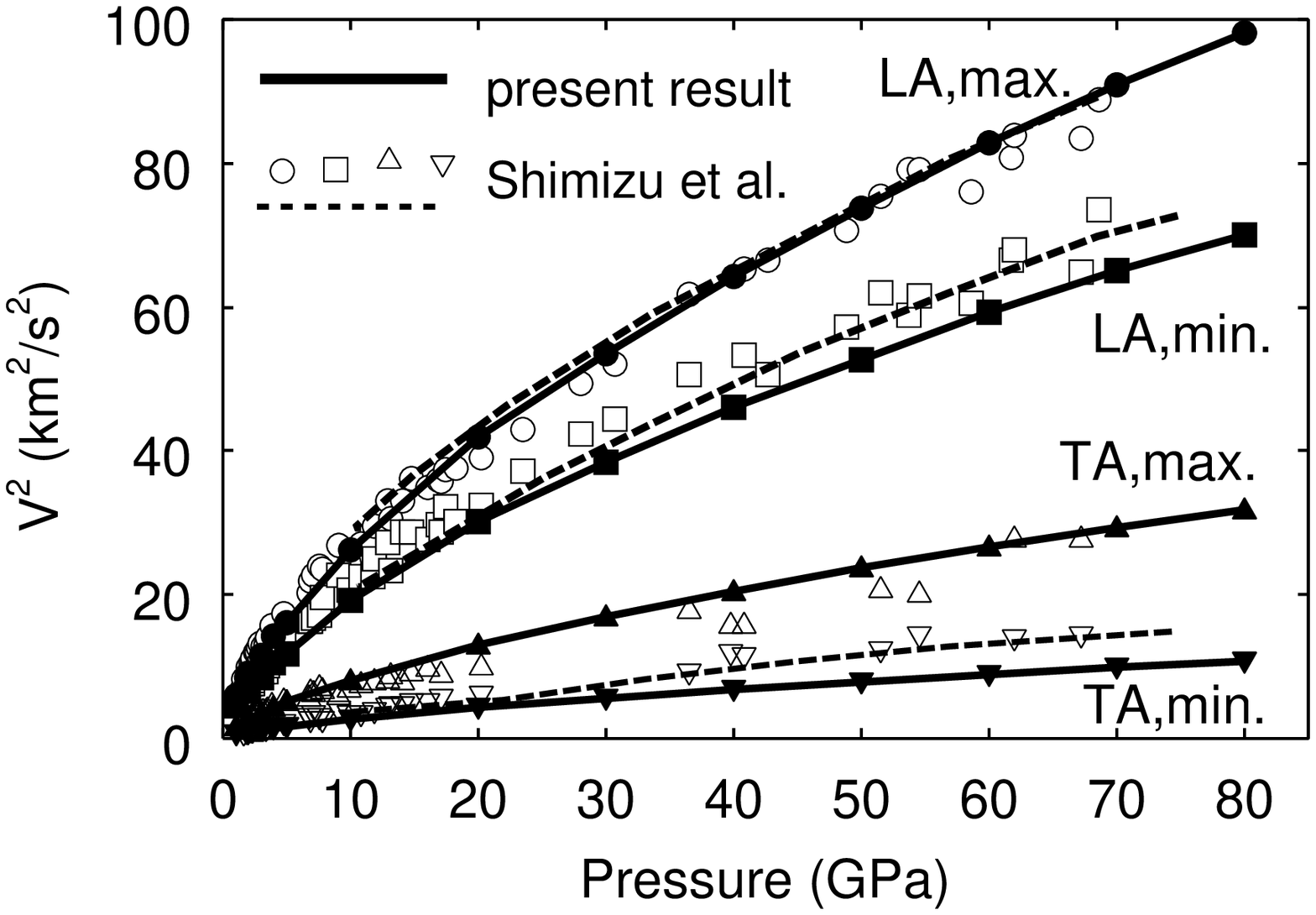}
\epsfxsize=\figwidth
\epsfbox{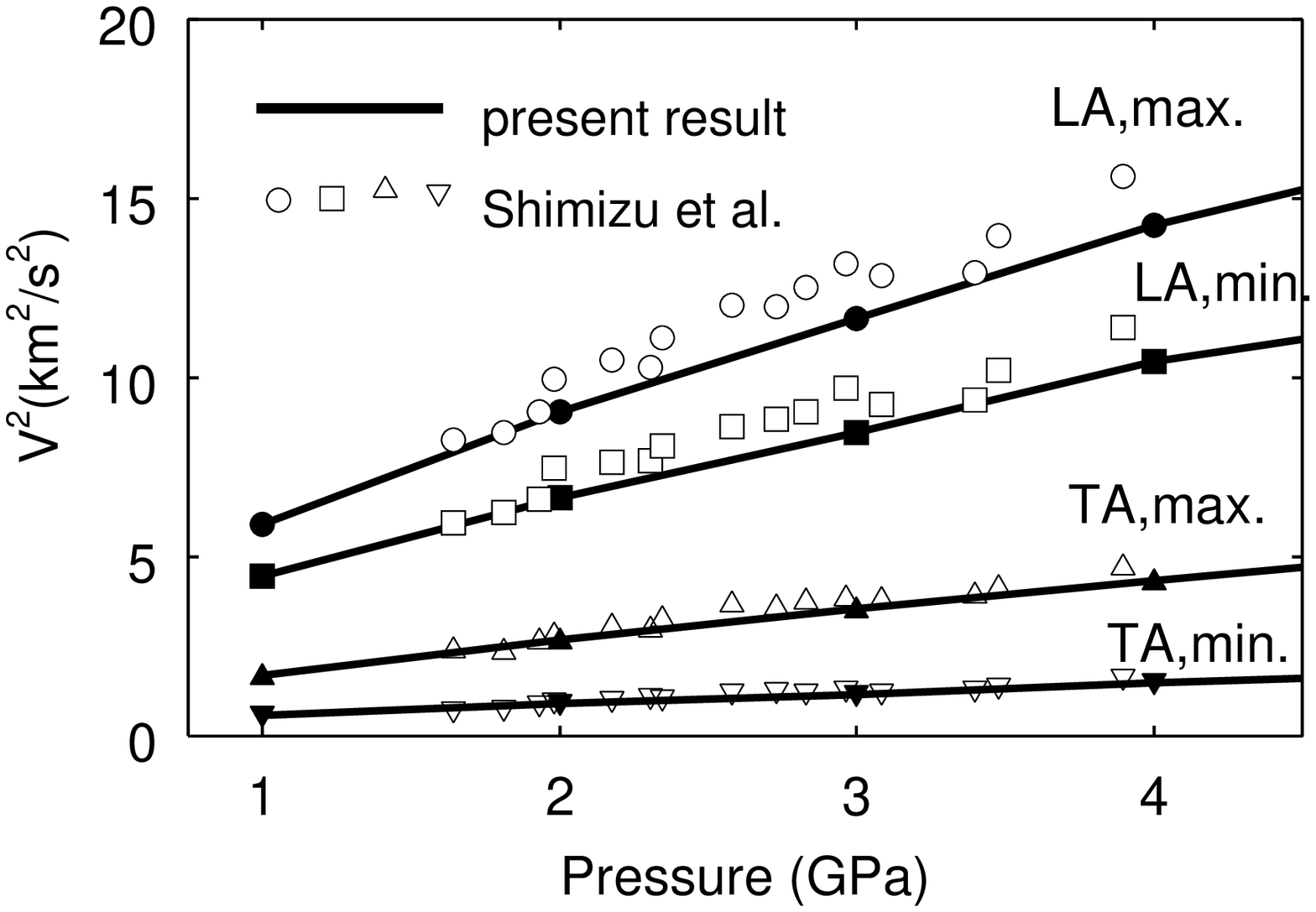}
\caption{
The pressure dependence of the squares of acoustic velocities, $v_{LA,max}^2$, $v_{LA,min}^2$, $v_{TA,max}^2$, and $v_{TA,min}^2$, of fcc solid argon. The solid curves with closed symbols represent the present result. Open symbols represent the experimental data obtained  by Shimizu et al.  \protect\cite{Shimizu2001}.
The envelope method was used for above 4 GPa and {\it in situ} Brillouin spectroscopy was used below 4 GPa.
The dashed curves are the envelope curves for the experimental data \protect\cite{Shimizu2001}.
}
\end{figure}

Figure~3 shows the pressure dependence of the squares of the acoustic velocities, which are related to the elastic stiffness coefficients by
\begin{eqnarray}
v_{LA,max}^2 &=& (c_{11} + 2c_{12} +4c_{44})/(3\rho) \\
v_{LA,min}^2 &=&  c_{11}/\rho \\
v_{TA,max}^2 &=&  c_{44}/\rho \\
v_{TA,min}^2 &=& (c_{11}-c_{12})/(2\rho)
.
\end{eqnarray}
The agreement between the present result and the experimental one becomes better than in Fig.~2, though theoretical $v_{TA,min}$ above 10 GPa seems slightly smaller than the experimental data. The difference between the present result and the experimental result in Fig.~2 may depend on how the experimental envelope curves are drawn in Fig.~3.


\begin{figure}
\epsfxsize=\figwidth
\epsfbox{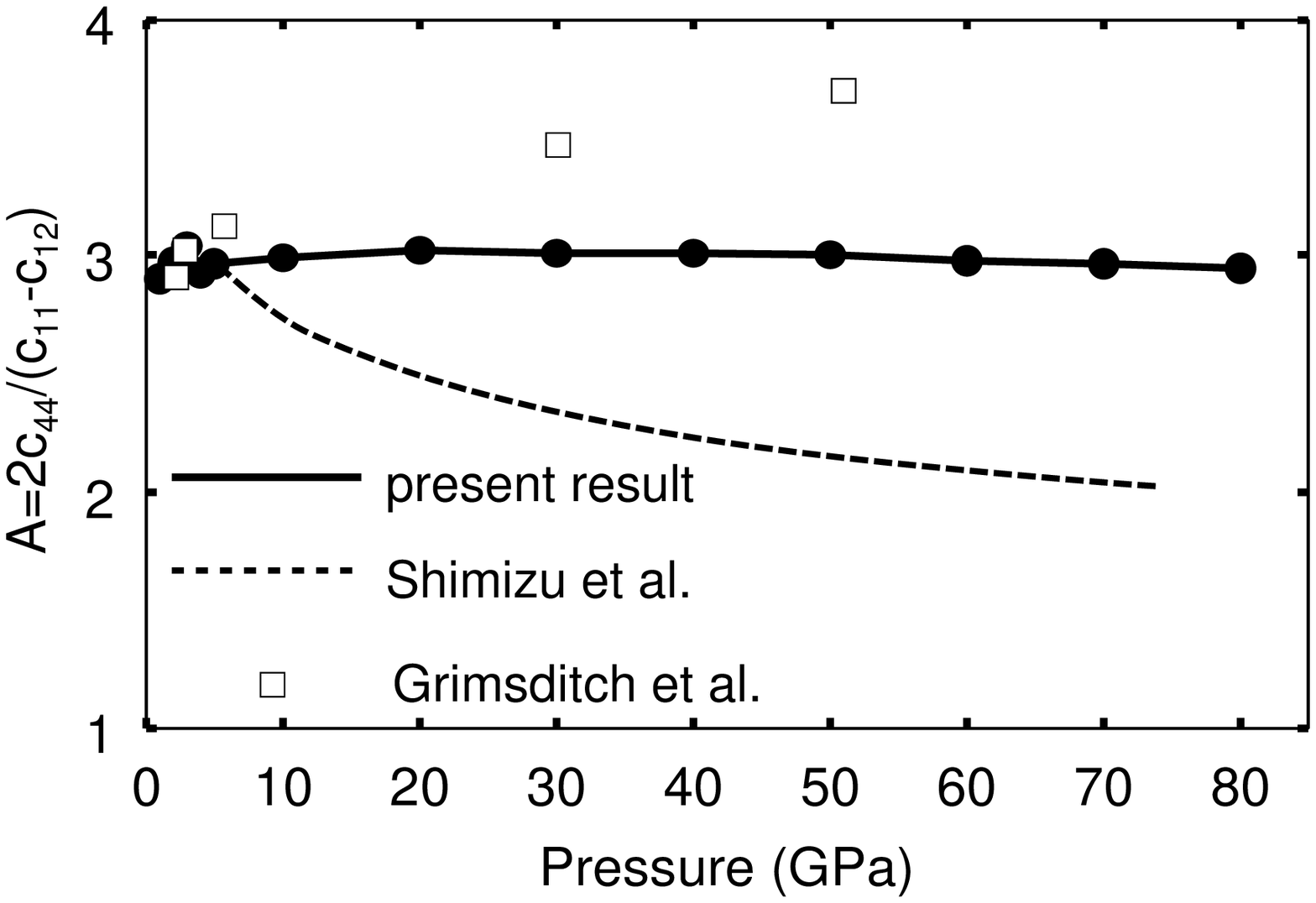}
\epsfxsize=\figwidth
\epsfbox{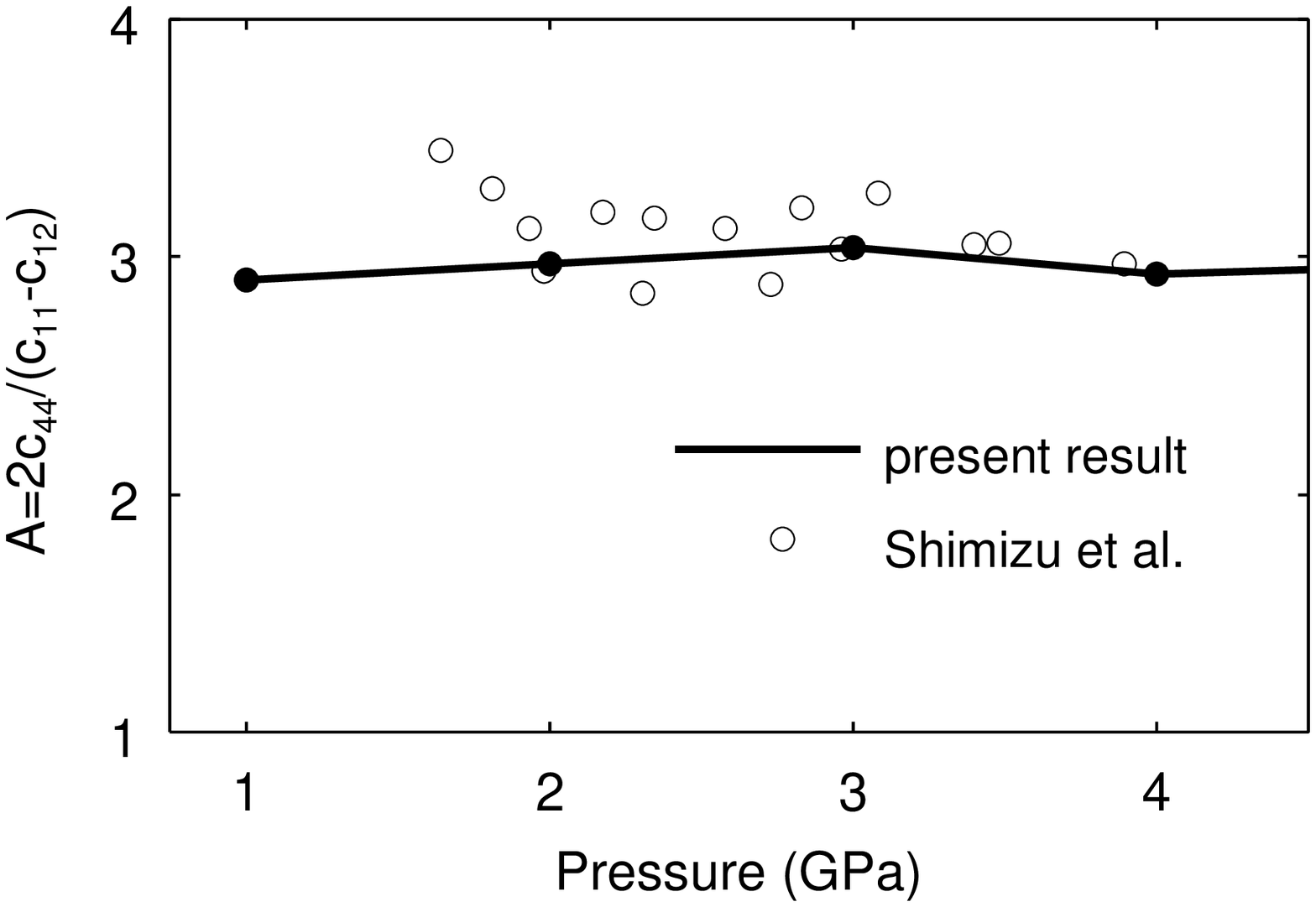}
\caption{
The pressure dependence of the elastic anisotropy, $A=2c_{44}/(c_{11}-c_{12})$, for fcc solid argon. 
Open squares represent the theoretical result by Grimsditch et al. \protect\cite{Grimsditch1986}.
Other symbols and lines has meaning as in Fig3.
}
\end{figure}

Figure~4 shows the pressure dependence of the elastic anisotropy $A=2c_{44}/(c_{11}-c_{12})$, which is the ratio of two shear moduli $c_{44}$ and $(c_{11}-c_{12})/2$, and which becomes unity for isotropic elasticity.
The anisotropy calculated between 1.6 and 4 GPa is approximately three, and agrees well with the experimental result. Above 4 GPa, the experimental anisotropy slowly decreases toward two, while the present result is almost constant up to 80 GPa.

\begin{figure}
\epsfxsize=\figwidth
\epsfbox{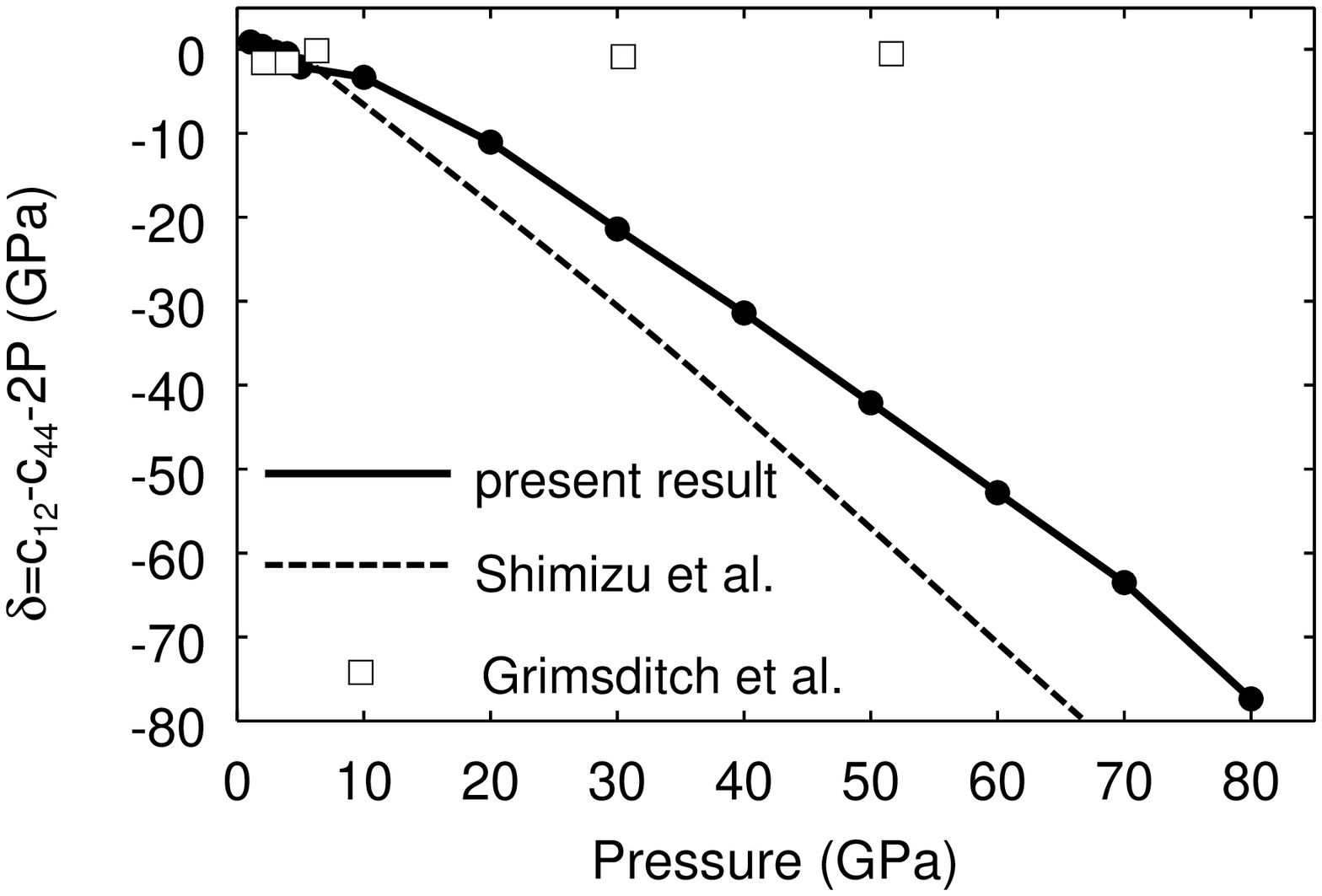}
\epsfxsize=\figwidth
\epsfbox{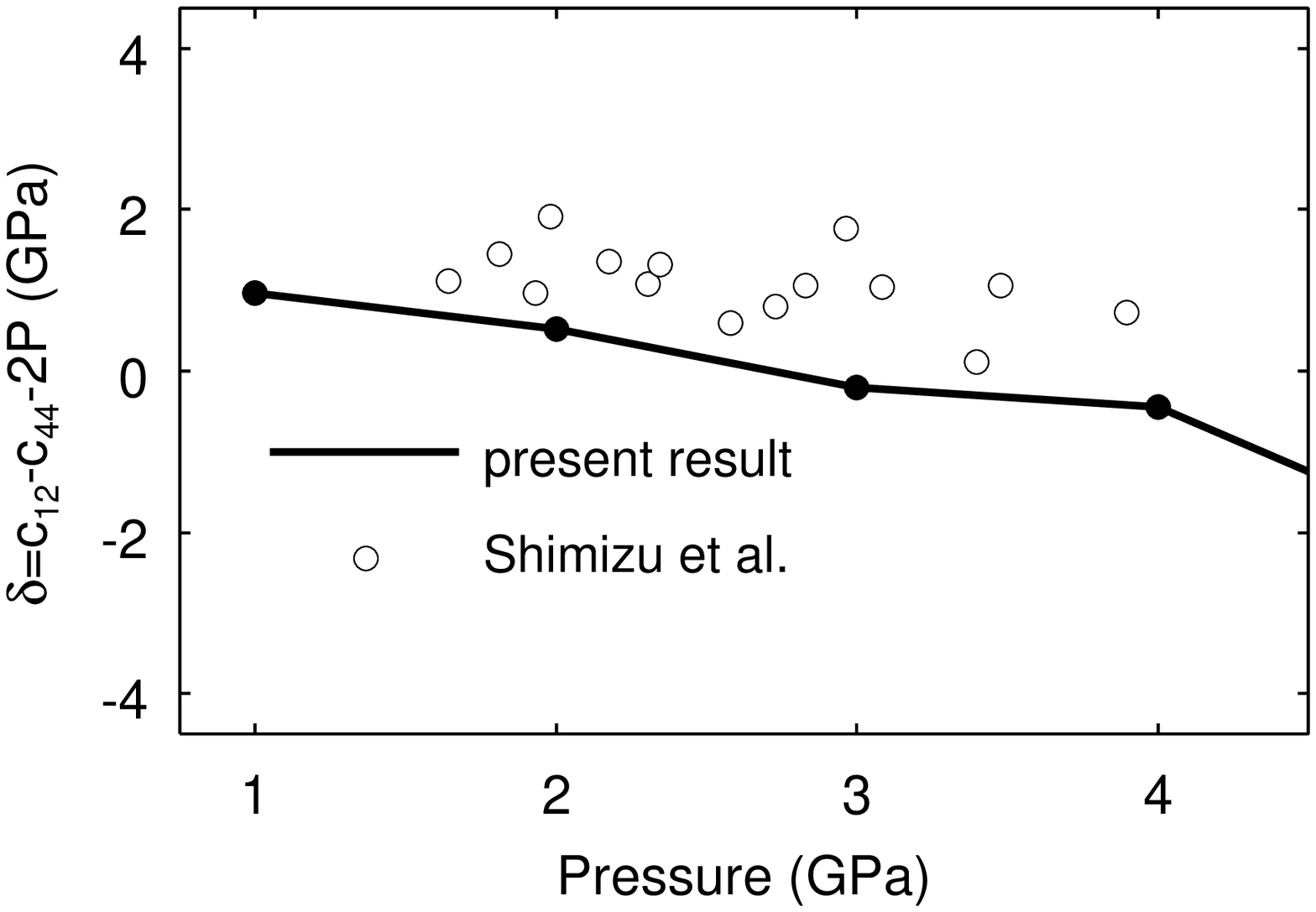}
\caption{
The pressure dependence of the deviation from the Cauchy relation, $\delta=c_{12}-c_{44}-2P$, for fcc solid argon.
Symbols and lines have meanings as in Fig.4.
}
\end{figure}

The deviation from the Cauchy relation $\delta=c_{12}-c_{44}-2P$ is a measure of the contribution from the non-central many-body force since the Cauchy relation $c_{12}=c_{44}+2P$ should be satisfied when interatomic potentials are purely central \cite{Grimsditch1986,Sinogeikin1999}.
Figure~5 shows the pressure dependence of $\delta $.
The deviation $\delta $ for the experimental result \cite{Shimizu2001} becomes larger as the pressure increases, which indicates that non-central many-body force becomes more and more important at high pressures.  The $\delta$ of the present result agrees well with the experimental result, indicating that first principles calculation with plane wave basis set and pseudopotentials can correctly describe the many-body force. The $\delta$ for the theoretical result of Grimsditch et al. is almost zero for all pressures since their theory is based on pair potentials.

In summary, we computed the elastic properties of fcc solid argon at high pressure by using first principles calculation at zero temperature with plane wave basis set, pseudopotentials and generalized gradient approximation for exchange-correlation interaction. We have shown that the standard density functional method with periodic boundary conditions at zero temperature and under constant pressures can be successfully applied for calculating elastic properties of rare gas solids at high pressures in contrast to those at low pressures where dispersion forces are important. Though the effects of thermal and zero-point vibrations were not included in the present calculation, thermal effects are expected to be small at high pressures. The long-range van der Waals' force is not also taken into account explicitly. However, the lattice constant and the elastic properties of solid argon at high pressures are mainly determined by the balance between the short-range repulsive force and the external pressure, and the van der Waals' force will be negligible.


The authors would like to thank G.~J.~Ackland and D.~M.~Bird for valuable discussions and providing CASTEP code, and H.~Shimizu and S.~Sasaki for their encouragement and providing their Brillouin spectroscopy data.
The research presented here was undertaken within a joint collaboration between the Computational Science Division of RIKEN (The Institute of Physical and Chemical Research, Wako-shi, Japan) and the United Kingdom Car-Parrinello (UKCP) consortium. The results presented here ware computed by using supercomputers at RIKEN, ISSP and NIG.

\clearpage


\begin{thebibliography}{99}

\bibitem[*]{tiitaka}
Corresponding author. \\
Email address: tiitaka@postman.riken.go.jp

\bibitem{Aziz1986}
R.~A.~Aziz, and M.~J.~Slaman,
Mol. Phys. {\bf 58}, 679 (1986).

\bibitem{Tse1998}
J.~S.~Tse, V.~P.~Shpakov, and V.~R.~Belosludov,
Phys. Rev. B {\bf 58}, 2365 (1998).

\bibitem{Lotrich1997}
V.~F.~Lotrich, and K.~Szalewicz,
Phys. Rev. Lett. {\bf 79}, 1301 (1997).

\bibitem{Rosciszewski2000} 
K.~Rosciszewski, B.~Paulus, P.~Fulde, H.~Stoll,, 
Phys. Rev. B {\bf 62}, 5482 (2000).

\bibitem{Shimizu1981}
H.~Shimizu, E.~M.~Brody, H.~K.~Mao and P.~M.~Bell,
Phys. Rev. Lett. {\bf 47}, 128 (1981).

\bibitem{Grimsditch1986}
M.~Grimsditch, P.~Loubeyre, and A.~Polian,
Phys. Rev. B {\bf 33}, 7192 (1986).

\bibitem{Shimizu2001} 
H.~Shimizu, H.~Tashiro, T.~Kume, and S.~Sasaki, 
Phys. Rev. Lett. {\bf 86}, 4568 (2001).

\bibitem{Shimizu1992}
H.~Shimizu, and S.~Sasaki,
Science {\bf 257}, 514 (1992).

\bibitem{Brazhkin2001}
V.~V.~Brazhkin and A.~G.~Lyapin,
JETP Letters, 73, 197 (2001).

\bibitem{Occelli2001}
F.~Occelli, M.~Krisch, P.~Loubeyre, F.~Sette, R.~Le~Toullec, C.~Masciovecchio, and J.-P.~Rueff,
Phys.~Rev.~B. {\bf 63}, 224306 (2001).

\bibitem{Kwon1995}
I.~Kwon, L.~A.~Collins, J.~D.~Kress, and N.~Troullier,
Phys. Rev. B {\bf 52}, 15165 (1995).

\bibitem{McMahan1986}
A.~K.~McMahan,
Phys. Rev. B {\bf 33}, 5344 (1986).

\bibitem{Payne1992}
M.~C.~Payne, M.~P.~Teter, D.~C.~Allan, T.~A.~Arias, J.~D.~Joannopoulos, 
Rev. Mod. Phys. {\bf 64}, 1045 (1992); 
CASTEP 4.2 academic version, licensed under the UKCP-MSI Agreement, 1999.

\bibitem{Vanderbilt1990}
D.~Vanderbilt, 
Phys. Rev. B {\bf 41}, 7892 (1990).

\bibitem{Perdew1996}
J.~P.~Perdew, K.~Burke, and M.~Ernzerhof,
Phys. Rev. Lett. {\bf 77}, 3865 (1996).

\bibitem{Monkhorst1976}
H.~J.~Monkhorst and J.~D.~Pack,
Phys. Rev. B {\bf 13}, 5188 (1976).

\bibitem{Francis1990} 
G.~P.~Francis and M.~C.~Payne, 
J. Phys. Condensed Matter {\bf 2}, 4395 (1990).

\bibitem{Wentzcovitch1995}
R.~M.~Wentzcovitch, N.~L.~Ross, and G.~D.~Price,
Phys. Earth Planet. Interiors {\bf 90}, 101 (1995).

\bibitem{Karki1997} 
B.~B.~Karki, L.~Stixrude, S.~J.~Clark, M.~C.~Warren, G.~J.~Ackland, and J.~Crain, 
Am.~Mineral. {\bf 82}, 51 (1997).


\bibitem{Wang1993}
J.~Wang, S.~Yip, S.~R.~Phillpot and D.~Wolf,
Phys. Rev. Lett. {\bf 71}, 4182 (1993).

\bibitem{Wang1995}
J.~Wang, J.~Li, S.~Yip, S.~Phillpot and D.~Wolf,
Phys. Rev. B {\bf 52} 12627 (1995).

\bibitem{Karki1997b} 
B.~B.~Karki, G.~J.~Ackland, and J.~Crain, 
J.~Phys. Condens. Matter {\bf 9}, 8579 (1997).


\bibitem{Sinogeikin1999}
S.~V.~Sinogeikin and J.~D.~Bass,
Phys. Rev. B {\bf59}, R14141 (1999).

\end{thebibliography}
\end{document}